\documentclass[vecphys]{svmult}

\usepackage{makeidx}         
\usepackage{graphicx}        
\usepackage{multicol}        
\usepackage[bottom]{footmisc}
\usepackage{natbib}
\usepackage{mesjournaux} 

\makeindex             

\begin{document}

\title*{MHD disc winds}
\author{Jonathan Ferreira}
\institute{Laboratoire d'Astrophysique de Grenoble, BP 53, 38041 Grenoble Cedex, France\\
Universit\'e Joseph Fourier, Grenoble, France\\
Jonathan.Ferreira@obs.ujf-grenoble.fr}
\maketitle

This is a doctorate level lecture on the physics of accretion discs driving magnetically self-confined jets, usually referred to in the literature as disc winds. 

I will first review the governing magnetohydrodynamic equations and then discuss their physical content. At that level, necessary conditions to drive jets from keplerian accretion discs can already be derived. These conditions are validated with self-similar calculations of accretion-ejection structures.

In a second part, I will critically discuss the biases introduced when using self-similarity as well as some other questions such as: Are these systems really unstable? Can a standard accretion disc provide the conditions to launch jets in its innermost parts? What is the difference between X-winds and disc-winds?
 
Finally, the magnetic interaction between a protostar and its circumstellar disc will be discussed with a focus on stellar spin down.

\section{The accretion-ejection paradigm}

This lecture is designed to be read with an accompanying file (pdf or ppt) where more illustrations and figures can be found. It can be retrieved at the URL: {\it  http://www-laog.obs.ujf-grenoble.fr/$\sim$ferreira/\-JETSET/school.html}. I also recommend the reviews of \citealt{koni00,ferr02}. In \citet{ferr06b}, a review of all MHD models for Young Stellar Objects has been made with a comparison of the corresponding jet kinetic observational properties. 

\subsection{A "universal" picture}

Actively accreting ''classical" T~Tauri stars (TTS) often display supersonic collimated jets on scales of a few 10-100~AU in low excitation optical forbidden lines. Molecular outflows observed in younger Class 0 and I sources may be powered by an inner unobserved "optical jet" (see Cabrit's contribution, this volume). These jet signatures are correlated with the infrared excess and accretion rate of the circumstellar disc \citep{cabr90,hart95}. It is therefore widely believed that the accretion process is essential to the observed jets, although the precise physical connection remains a matter of debate: do the jets emanate from the star, the circumstellar disc or the magnetospheric star-disc interaction?

One argument in favor of accretion-powered disc winds is its "universality" \citep{livi97}. Indeed, self-collimated jet production from accretion discs is also invoked to explain an accretion-ejection correlation observed in compact objects (i.e. some active galactic nuclei, quasars and X-ray binaries, see eg. \citealt{merl03} and references therein). The underlying idea is quite simple: accretion discs around a central object can, under certain circumstances and whatever the nature of this object, drive jets through the action of large scale magnetic fields. These fields would tap the mechanical energy released by the mass accreting in the disc and transfer it to the fraction that is ejected \citep{blan82}. The smaller the fraction and the larger the final jet velocity. One thing that must be understood is how the presence of such jets modifies the nature of the underlying accretion flow. Many papers in the literature actually assume (implicitly or not) that the accretion disc resembles a standard accretion disc as envisioned by \citet{shak73,fran02}. This is wrong as will be shown later. 

A Magnetized Accretion-Ejection Structure (hereafter MAES)\index{MAES}\index{wind models, disc wind, extended} is an accretion disc where accretion and ejection are interdependent processes. As such, it is composed of an accretion disc (called hereafter JED for Jet Emitting Disc) thread by a large scale magnetic field of bipolar topology and giving rise to the two bipolar jets. The goal of the study of a MAES is to obtain\\ 
(1)- the conditions allowing for a steady state accretion-ejection process;\\
(2)- the ejection to accretion rates ratio as function of the disc physical conditions;\\
(3)- the jet properties (kinematics, power, shape) as function of the disc properties.

\subsection{From magnetostatics to magnetohydrodynamics}

Magnetohydrodynamics (MHD) is the theoretical framework required to describe the interaction between an ionized gas and magnetic fields. But magnetostatics is very helpful to understand basic mechanisms.
  
\begin{figure}
\centering
\includegraphics[height=4cm]{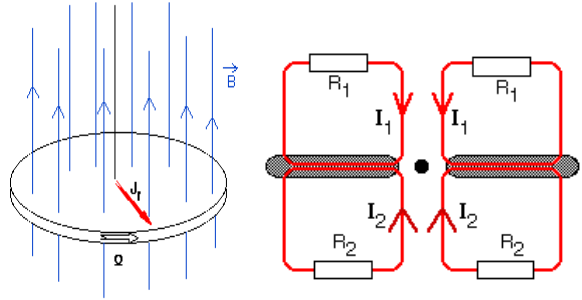}
\caption{Left: A rotating disc embedded in a magnetic field induces a current leading to a magnetic braking (Barlow's wheel: see e.g. {\it http://www.sparkmuseum.com/MOTORS.HTM} for many illustrations). Right: A MAES can be seen as two independent electric circuits, each corresponding to a jet. Asymmetric jets can thus be easily achieved, even with a symmetric poloidal field.}
\label{fig:barlow}
\end{figure}

A zeroth order description of a MAES is that of a rotating conducting disc thread by a magnetic field aligned with the rotation axis (much alike a  Barlow's wheel\index{Barlow's wheel}). According to Faraday's induction law, an electromotive force (emf) across the disc, $e= \int (\vec u \wedge \vec B) \cdot \vec dr = \int \Omega r B_z dr$,  creates an electric potential difference between the disc center and its border (Fig.~\ref{fig:barlow}). If some conducting wire connects the border to the center, closing thereby an electric circuit, then a radial electric current is induced. Because of this current $I$, the disc becomes prone to a Laplace force, $F = \int I B_z dr$, which will slow down the disc (Lenz's law). One could also say that the field "resists" to the shear provoked by the rotation (the current $I$ induces a toroidal component $B_\phi$). But such a "mechanical" view of the magnetic field disregards its electromagnetic nature and one may tend to forget that electric currents must be maintained and able to flow...

In astrophysics, the disc is made of gas that, provided it can cross the field lines, will accrete towards the central object as it looses angular momentum. This angular momentum is linked to the electric current flowing in the jets: the jet kinetic power is fed by the flux of magnetic energy provided at the disc surface. Note that while the streamlines of the ejected material go to infinity those of the current density must be closed and return to the disc where the emf is.

This is actually the reason why jet collimation\index{collimation, magnetic} is a subtle issue \citep{heyv89b,heyv03,okam03}. Make a cut at a distance $z$ of a jet and compute the total current\index{electric current} flowing inside it, namely $I= \int dr 2\pi r J_z$. If this current is non zero and (for instance) negative, then one might say that the Laplace force will be directed towards the jet axis (Ampere's theorem tells that $B_\phi$ is negative in that case). This is the basic idea of the "magnetic hoop stress" that provides a self-confinement to jets\index{magnetic, hoop stress}. However, the local magnetic force is actually $\vec J \wedge \vec B$ and depends on the radial distribution of $J_z(r)$! This depends on the lateral boundary conditions (jet axis and outer edge) but also on what happened upstream (or in the past, if we follow a lagrangean particle): since jet acceleration is a conversion of electric into kinetic power, then jet collimation depends as much on jet acceleration. One cannot therefore solve the jet problem assuming for instance the shape of the field lines: the full MHD equations must be solved.  

The current flowing inside jets is precisely the current that allows for accretion. The accretion-ejection phenomenon has therefore to be viewed as a global electric current system\index{electric current}.

\subsection{Basic assumptions}

Modeling a MAES requires several assumptions:

{\bf (1) Presence of a large scale vertical magnetic field} in the disc. Its origin and amplitude remain an open question. For the purpose of illustration, we will assume a positive vertical component $B_z$ anchored in the disc (bipolar topology)\index{magnetic field, bipolar}.

{\bf (2) Single-fluid MHD}: matter is assumed ionized enough so that all species (ions, neutrals and electrons) are well coupled and can be treated as a single fluid. Such an assumption should always be verified a posteriori for any model but it is seldom made (see e.g. \citealt{garc01b} for how to do it).

\begin{figure}[t]
\centering
\includegraphics[height=4cm]{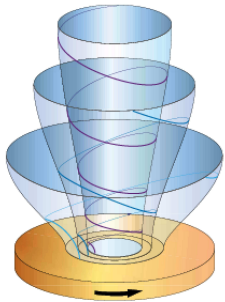}
\caption{Axisymmetric jets are made of magnetic surfaces of constant magnetic flux nested around each other and anchored in the disc\index{magnetic, surface}. Each surface behaves like a funnel whose shape depends on the transfield equilibrium. Solving the jet equations requires to specify several quantities (see text).}
\label{fig:model}
\end{figure}

{\bf (3) Axisymmetry}: using cylindrical coordinates ($r$, $\phi$, $z$) all quantities are assumed to be independent on $\phi$, the jet axis being the vertical axis. Then, $E_\phi =0$ and all quantities can be decomposed into poloidal (the ($r$, $z$) plane) and toroidal components, eg. $\vec u =
\vec{u_p} + \Omega r \vec{e_\phi}$ and $\vec B= \vec{B_p} + B_{\phi} \vec{e_\phi}$. A bipolar magnetic configuration can then be described with $\vec{B_p} = \frac{1}{r} \nabla a \wedge \vec{e_\phi}$, where the magnetic flux function $a (r,z)$ is an even function of $z$ and with an odd toroidal field $B_\phi(r,-z)= -B_\phi(r,z)$ (Fig.~\ref{fig:model}).

{\bf (4) Non-relativistic MHD}, since observed motions are non-relativistic (this criterion is enough as long as MHD ordering applies).

{\bf (5) Steady-state}: all astrophysical jets display proper motions and/or emission nodules, showing that they are either prone to some instabilities or that ejection is an intermittent process. However, the
time scales involved in all objects (from 1 to $10^2$ yrs) are always larger than the orbit time scales in the innermost regions of the underlying accretion disc (close to the star). Therefore, a steady
state approach is appropriate as a first step, while numerical simulations will be required to investigate time-dependent flows. 

\subsection{Governing MHD equations}

According to the aforementioned assumptions, we use the following set of MHD equations (in MKSA units):

\noindent {\bf Mass conservation}
\begin{equation} 
\nabla \cdot \rho \vec u  = 0
\label{eq:1}
\end{equation}
\noindent {\bf Momentum conservation}\index{MHD equations, momentum}
\begin{equation} 
\rho \vec u \cdot \nabla \vec u = -\nabla P - \rho \nabla \Phi_G 
+ \vec J \wedge \vec B  +  \nabla \cdot {\mathsf T}
\label{eq:2}
\end{equation}
where $\Phi_G$ is the central star gravitational potential and $\vec J= \nabla \wedge \vec B/\mu_o $ is  the electric current density. The last term (with the stress tensor ${\mathsf T}$)  is actually due to a sustained turbulence\index{angular momentum transport, turbulent} inside the disc (it vanishes outside) which allows to transport angular momentum radially in the outward direction (see Terquem's contribution). It is presumably due to the presence of small scale magnetic fields but is usually grossly modeled by an anomalous viscosity\index{transport coefficients, anomalous viscosity} $\nu_v= \alpha_v C_s h$, where  $\alpha_v$ is a free parameter\index{disc, alpha (Shakura-Sunyaev)}, $C_s$ the disc sound speed and $h(r)$ the local disc vertical scale height \citep{shak73,fran02}.

\noindent {\bf Ohm's law}\index{MHD equations, Ohm's law} and {\bf toroidal field induction}\footnote{See Pelletier's contribution in this volume. Remember that $E_\phi=0$ while some algebra is required in order to derive Eq.~\ref{eq:4} from the induction equation.}\index{MHD equations, induction}
\begin{eqnarray}
  \eta_m \vec J_{\phi}  &=& \vec{u_p} \wedge \vec{B_p}\label{eq:3}\\
  \nabla \cdot (\frac{\nu'_m}{r^2}\nabla rB_{\phi}) & = &
  \nabla \cdot \frac{1}{r}(B_{\phi} \vec{u_p} - \vec{B_p}\Omega r)
  \label{eq:4} 
\end{eqnarray}
where $\eta_m=\mu_o \nu_m$ and $\eta'_m= \mu_o\nu'_m$ are anomalous magnetic  
resistivities\index{transport coefficients, anomalous resistivity}. The origin of these resistivities is the same as for viscosity, namely turbulence and they also vanish outside the disc. One expects turbulent  media to display anomalous transport effects of heat, momentum but also magnetic flux. Note however that rotation in a Keplerian accretion disc introduces a strong dynamical constraint. Indeed, the shear induced by rotation will unavoidably lead to huge toroidal magnetic fields until reconnection\index{reconnection} takes place (triggered by e.g. the tearing mode instability\index{instabilities, tearing}). As a consequence the amount of magnetic dissipation in the toroidal direction might be much larger than in the poloidal direction. This has lead to the introduction of two anomalous coefficients,  $\eta_m$ ($\nu_m$) and  $\eta'_m$ ($\nu'_m$), related respectively to the poloidal and toroidal fields \citep{ferr95}.

\noindent {\bf Perfect gas law}\index{MHD equations, state}
\begin{equation}
P= \rho \frac{k_B}{\bar \mu m_p} T
\label{eq:5}
\end{equation}
where $m_p$ is the proton mass and $\bar \mu$ a generalized "mean molecular
weight" (in a fully ionized plasma $\bar \mu= 1/2$). This expression assumes that all fluids (electrons, neutrals and ions) have the same temperature $T$. This is fulfilled only if the thermalization time scale (usually done through collisions) is short enough. Such an assumption should always be verified a posteriori \citep{garc01b}.

\noindent {\bf Energy equation}\index{MHD equations, energy}
\begin{equation}
P \vec u \cdot \nabla \ln \frac{T}{\bar \mu} = (\gamma-1) \left ( Q + P  \vec u \cdot \nabla \ln \rho \right ) 
\label{eq:6}
\end{equation}
where $Q= Q^+ - Q^-$ is the sum of all heating $Q^+$ and cooling $Q^-$ terms (including thermal conduction) and $\gamma$ the adiabatic index. There are many unsolved issues related to this exact equation for a single fluid.\\ 
(1) Inside the disc, turbulence leads to an energy dissipation $Q^+ = \eta_m J^2_{\phi}\ +\ \eta'_mJ^2_p + \rho \nu_v | r \nabla \Omega |^2$, respectively Joule and "viscous" heating\index{heating, Joule}\index{heating, turbulent}\index{heating, viscous}, but also to a cooling due to an energy transport by anomalous thermal conductivity\index{transport coefficients, anomalous thermal conductivity}. Moreover, the disc being optically thick, the radiation transport critically depends on the local opacity regime, which varies both with radius and height. Moreover, the disc surface is also the optically thick-thin transition, which is always an issue (see \citealt{ferr95} for a discussion). Besides, the energy equation in a standard accretion disc is usually written $Q^+=Q^-$, the other terms being of the order $(h/r)^2$ \citep{fran02}. But these terms are important in the jet and cannot be neglected. \\
(2) In the jet itself, although radiation may not be the dominant cooling term, it must be taken into account  if one desires to compute e.g. the jet (forbidden or permitted) emission lines or even radio continuum.  A realistic and self-consistent treatment of the energy equation is therefore still out of range (even if one decouples the disc and its jets) and some stratagems must be used. 

The simplest way to deal with the energy equation in a MAES (valid in both the disc and its jets) is to use a polytropic equation of state\index{MHD equations, state} $P= K \rho^{\Gamma}$, where the polytropic index $\Gamma$ can be set to vary between 1 (isothermal case) and $\gamma$ (adiabatic case). Note that $K$ has to vary radially but remains constant along each field line: the jet entropy is thus fixed by the conditions prevailing at the disc surface. 

A more sophisticated approach can be done by prescribing the function $Q$ along the field lines (this is equivalent to prescribing a variation of the polytropic index $\Gamma$). This will be discussed further in Section~3.

\section{Physics of Jet Emitting Discs}

In this section, I will briefly discuss all the relevant physical effects that have to be covered in order to consistently describe Jet Emitting Discs or JEDs.\index{disc, jet emitting disc (JED)}

\begin{figure}
\centering
\includegraphics[height=3cm]{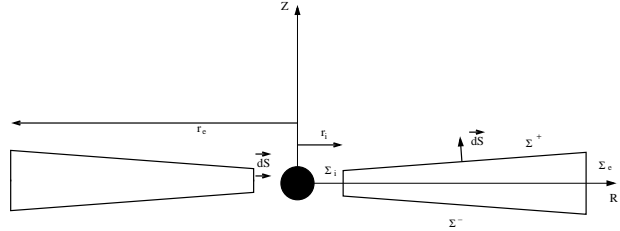}
\caption{Sketch of the Jet Emitting Disc (JED) established between $r_i$ and $r_e$. The surface of the jet is determined by the magnetic surface anchored on $r_e$. While the inner radius $r_i$ is probably defined by some equilibrium with the stellar magnetosphere (see Section~4.6), the outer radius $r_e$ is free (it depends mostly on the magnetic flux available in the disc).}
\label{fig:volume}
\end{figure}

\subsection{Mass conservation}

The disc accretion rate is defined as $\dot M_a = -2 \int_0^h 2 \pi r \rho u_r dz$. In a standard accretion disc (hereafter SAD) $\dot M_a$ is a constant both in time and radius\index{accretion, rate}. On the contrary, a JED displays mass loss at its surfaces so that $\dot M_a$ must vary with the radius. This mass loss is parametrized by $\dot M_a(r) \propto r^\xi$ where the ejection index $\xi >0$ is a measure of the disc ejection efficiency\index{disc, ejection index}:  the larger $\xi$ the larger the mass loss. The global mass conservation in a JED is $\dot M_a(r_e) - 2 \dot M_j = \dot M_a(r_i)$ where $r_e$ and $r_i$ are respectively the outer and inner radii of the JED and $\dot M_j$ is the mass flux\index{ejection, rate} from one side of the disc. The ejection to accretion rates ratio\index{ejection, efficiency} is $2 \dot M_j/\dot M_a(r_e) \simeq \xi \ln (r_J/r_i)$ and depends on both the ejection index $\xi$ and the radial extent of the JED (it will be shown later that $\xi$ is smaller than unity). The goal is of course to compute $\xi$ as a function of the disc physical conditions.

\subsection{Poloidal field diffusion}

Let us assume a smooth flux function $a$ (see Sect.~1.3) so that $a(r,z) \simeq a_o(r) (1 - z^2/2l^2)$ where $l(r)$ is the magnetic flux vertical scale height. Then, the bending of the poloidal field lines is measured at the equatorial plane by the magnetic Reynolds number\index{Reynolds, magnetic number} ${\cal R}_m = -ru_r/\nu_m = r^2/l^2$. Such a bending is due to the interplay between advection by the accreting material and the turbulent magnetic diffusivity $\nu_m$. It has been prescribed with $\nu_m= \alpha_m V_A h$, where $V_A= B_z/\sqrt{\mu_o \rho}$ is the Alfv\'en speed at the disc midplane \citep{ferr93a}. 
 
Now, magneto-centrifugal acceleration\index{ejection mechanism, magneto-centrifugal} requires field lines bent enough at the disc surface, namely $B_r^+ \geq B_z$ (\citealt{blan82}, quantities evaluated at $z=h$ are denoted with a superscript "+"). Since $B_r^+/B_z \simeq {\cal R}_m h/r$ this implies  ${\cal R}_m \geq r/h$.

\subsection{Angular momentum conservation}

The disc angular momentum can be transported by two means: (a) radially through a "viscous" turbulent torque  which is probably triggered and sustained by an MHD instability such as the magneto-rotational instability (see Terquem's contribution, \citealt{balb03} and references therein); (b) vertically by the jets\index{angular momentum transport, jet}. The viscous torque writes  $F_{visc,\phi} \sim - \alpha_v P/r$ where $P$ is the total (gas+radiation) pressure and $\alpha_v$ the so-called Shakura-Sunyaev parameter\index{disc, alpha (Shakura-Sunyaev)}. The torque\index{magnetic, torque} due to the jets writes $F_{mag,\phi}= J_zB_r - J_rB_z$ and its vertical behavior strongly depends on the radial current density $J_r$. At the disc midplane  $F_{mag,\phi}=- J_r B_z \sim B_\phi^+B_z/\mu_oh$ and the disc angular  momentum conservation reads
\begin{equation}
1 + \Lambda \simeq  - \frac{r u_r}{\nu_v} = {\cal R}_e =    {\cal R}_m \left ( \frac{\nu_m}{\nu_v} \right ) 
 \end{equation}
 where $\nu_v$ is the turbulent "viscosity" and $\Lambda= F_{mag,\phi}/F_{visc,\phi}$. In a turbulent medium,  one usually assumes that all anomalous transport coefficients are of the same magnitude so that $\nu_m \sim \nu_v$. In that situation, one gets the following consequences:\\
 - In a SAD\index{disc, standard}, there is no jets and $\Lambda =0$. Then ${\cal R}_m \simeq  {\cal R}_e \sim 1$ and, indeed,  field lines are too straight for a magneto-centrifugal driving \citep{lubo94a};\\
  - In a JED, jets require ${\cal R}_m \sim {\cal R}_e \geq r/h$ and thus $\Lambda \sim r/h \gg 1$: all the angular momentum must then be carried away by the jets, which results in an accretion velocity much larger than in a SAD. The "viscous" torque is totally negligible (in contrast to what is often assumed, e.g. \citealt{ogil98}).  
  
This very important constraint ($\Lambda \geq r/h$) can only be achieved if $- B_\phi^+ B_z/\mu_o \sim P$, that is with equipartition fields \citep{ferr95}.  Let us introduce here two important parameters: the disc magnetization\index{disc, magnetization} $\mu = B_z^2/\mu_o P$ and the magnetic shear $q \simeq - B_\phi^+/B_z$.  If $q\mu$ is not close to unity then no magneto-centrifugally driven jets can be launched from accretion discs.      

\subsection{Toroidal field induction}

Magnetic driving of jets requires that the magnetic field starts to accelerate material at the disc surface. Hence, a JED must provide a transition from $F_{mag,\phi}<0$ at $z=0$ to $F_{mag,\phi}>0$ at $z=h$ and beyond. The only way to achieve this is by allowing $J_r$ to decrease on a disc scale height. The vertical profile of $J_r$ is provided by the induction equation (\ref{eq:4}) which, in a thin accretion disc, writes (see \citealt{ferr95})
\begin{equation}
\eta_m' J_r \simeq \eta_o' J_o + r \int_0^z dz\, \vec B_p \cdot \nabla \Omega - B_\phi u_z 
\end{equation}
where $\eta_o' J_o = \eta_m' J_r (z=0)$. With no differential rotation, $J_r$ would remain constant and so would $F_{mag,\phi}$ ($<0$). In order to make $J_r$ decrease on a disc scale height, the disc differential rotation term must balance the term $\eta_o' J_o$, due to the Faraday's induction law (the Barlow's wheel current). This can be done as long as $- B_\phi^+/B_r^+ \sim 1/\alpha_m$ (with $\nu_m \sim \nu_m'$). Using the fact that $B_r^+ \geq B_z$, one gets a magnetic shear $q \geq 1/\alpha_m$. 
 
\subsection{Disc vertical equilibrium}

It is worthwhile to consider the following general equality
\begin{equation}
(\vec J \wedge \vec B_p) \cdot \vec B_\phi = - (\vec J_p \wedge \vec B_\phi) \cdot \vec B_p
\end{equation}
When the magnetic torque (lhs) is negative, so must be the projection of the Lorentz force on the poloidal field (rhs). Thus, deep within the disc, the poloidal Lorentz force is directed outwardly and towards the disc midplane. A quasi MHS equilibrium is therefore established with the balance between the total (gas+radiation) pressure gradient on one side and the magnetic compression due to the radial and toroidal field components and the gravity on the other side.  Now, as one goes up in $z$ and the magnetic torque changes its sign, the disc material starts to be azimuthaly accelerated. Correspondingly, the projection of the Lorentz force becomes also positive and helps to lift material out of the disc.  

\begin{figure}
\centering
\includegraphics[height=3cm]{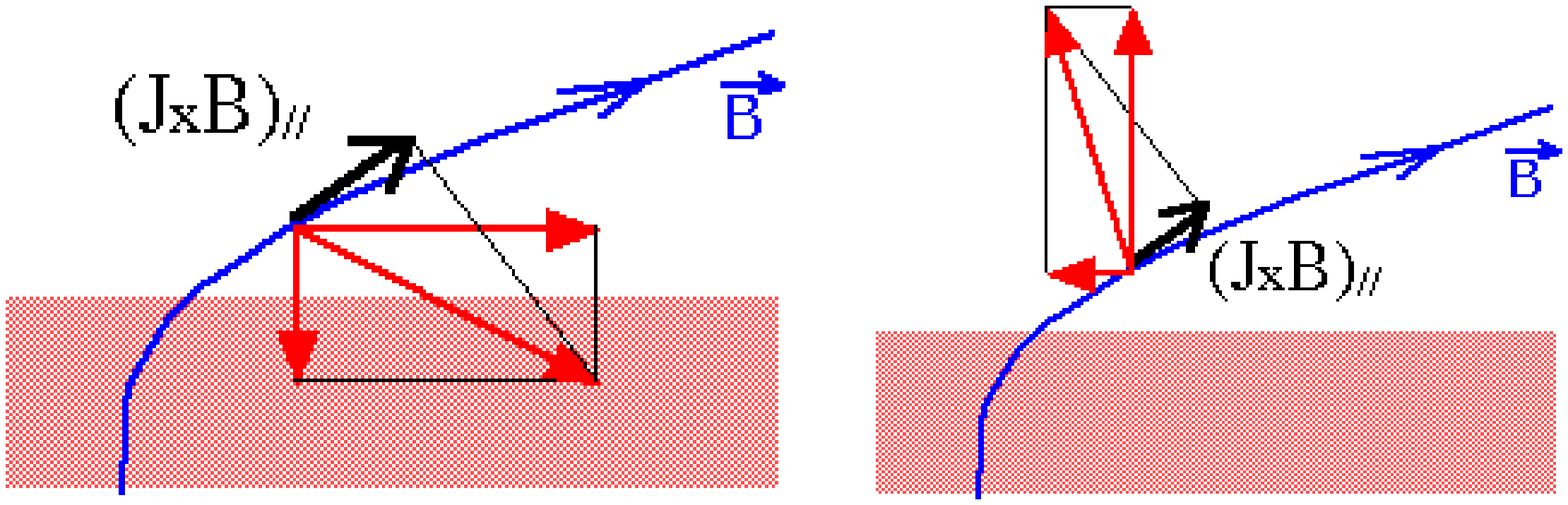}
\caption{Magnetic acceleration\index{ejection mechanism, magneto-centrifugal} arises whenever the projection of the Lorentz force on a poloidal field line becomes positive. This can be achieved in two ways, either with a downward vertical magnetic compression or a strong outward pressure force due to the toroidal field. The former leads to a small ejection efficiency and has current\index{electric current} lines coming out of the disc surface ($J_z >0$) and entering at the inner radius. The latter has a strong ejection efficiency with the current entering the disc at its surfaces ($J_z <0$). Only small ejection efficiencies allow for steady state solutions \citep{ferr97}.}
\label{fig:2courants}
\end{figure}

This can be done in two ways (Fig.~\ref{fig:2courants}): (a) with a negative vertical component of the Lorentz force but a large radial component; (b) with a positive vertical component and a smaller negative radial component. Case (a) corresponds to a small mass flux ($\xi < 1/2$) where disc material must be lifted against the magnetic compression by the sole effect of the (gas+radiation) pressure gradient. Case (b) leads to a large mass flux ($\xi> 1/2$) because of the magnetic pull due to the toroidal field pressure.

In fact, it can be shown analytically that only solutions with $\xi< 1/2$ can be stationary: solutions with large mass fluxes do not have enough power to allow for super-Alfv\'enic jets \citep{ferr97}. This has an important consequence on disc physics. Since $B_r^+ \geq B_z$, the total pressure gradient can overcome the magnetic compression due to $B_r$ only if $\mu$ is not larger than unity. The same constraint holds for the toroidal field which implies that $\alpha_m$ must be of the order of unity. Finally, using the fact that $\Lambda \sim r/h \gg 1$ in a JED, one obtains that $\mu$ cannot be too small. Thus, the parameter space for a JED is $\mu \sim q \sim \alpha_m \sim 1$. Note also that a JED is thinner than a SAD because of the additional magnetic compression \citep{ward93,ferr95}. 
 
\subsection{Disc radial equilibrium}

The quasi MHS radial equilibrium leads to an angular velocity
\begin{equation} 
\Omega^2 = \Omega_K^2 \left ( 1 + \frac{\partial P/\partial r}{\rho \Omega_K^2 r} 
- \frac{(\vec J\wedge \vec B)_r}{\rho \Omega_K^2 r}  + \frac{u_r^2}{\Omega_K^2 r^2} 
\frac{\partial \ln u_r}{\partial \ln r}\right )
\end{equation}
The deviation to the Keplerian rotation law $\Omega_K= \sqrt{GM/r^3}$ due to the radial (gas+radiation) pressure gradient is roughly of order $(h/r)^2$ at each altitude. This is because $P$ scales with the density, which is not the case of the radial magnetic tension. At the disc midplane, it causes a deviation which is of the order $\sim \mu {\cal R}_m (h/r)^2 \sim h/r$ but increasing as $1/\rho$.
Thus, thin accretion discs with $h/r \ll 1$ will be mostly rotating at (sub-) Keplerian speeds\index{velocity, Keplerian} but a problem arises when $h\sim r$ (as in ADAFs \citealt{nara98} or in self-gravitating discs, for instance). Indeed, it would imply a negative rhs at the disc surface which certainly means that no steady-state accretion-ejection solution can be found in that case. Note however that there is a priori no reason to ever have $h \sim r$ in JEDs: they are colder (see below) and more squeezed (see above) than a SAD.

\subsection{Energy budget}

The global energy budget is obtained by applying the energy conservation equation to the whole volume occupied by the JED. This equation writes $P_{acc} = 2 P_{MHD} + 2 P_{rad}$ where 
\begin{equation}
P_{acc} \simeq \frac{GM \dot M_a(r_e)}{2 r_i}
\end{equation} 
is the mechanical power\index{accretion, luminosity} liberated by the accreting material between $r_e$ and $r_i$, $P_{MHD}= \int \vec S_{MHD} \cdot \vec dS$ is the flux through one disc surface of the MHD Poynting\index{Poynting flux} vector $\vec S_{MHD} = \vec E \wedge \vec B/\mu_o \simeq - \Omega_K r B_\phi \vec B_p/\mu_o$. So, energy conservation in a JED tells us that the available accretion power is shared between a flux of electromagnetic energy powering the jets and radiation due to heat dissipation within the disc ($2P_{rad}= P_{diss}$). This dissipation is due to the fact that in a disc where turbulent  magnetic diffusivity/resistivity and viscosity are assumed, there is always some heat production. The simplest and crudest way to estimate this dissipation is using effective transport coefficients so that it writes $P_{diss} = \int_V dV (\eta_m J_\phi^2 + \eta_m' J_p^2 + \eta_v (r \partial \Omega/\partial r)^2)$, namely Joule and "viscous" heating\index{heating, Joule}\index{heating, viscous}.

Since the "viscous" torque is negligible with respect to the jet torque, only a small fraction of the energy will be dissipated by viscosity. On the other hand, the most interesting aspect of angular momentum removal by jets is that the associated Joule dissipation implies also only a small fraction of the available energy. As a consequence, most of the liberated accretion power goes into the jets \citep{ferr93a,ferr95}! \index{jet, mechanical power} Precisely, this can be written as
\begin{equation}
\frac{2P_{MHD}}{P_{acc}} \simeq \frac{\Lambda}{1 + \Lambda} \, \, \, \, \, \, \, \, \mbox{      and      } \, \, \, \, \, \, \, \,
\frac{2P_{rad}}{P_{acc}} \simeq \frac{1}{1 + \Lambda} 
\end{equation}
where the ratio of the jet to the viscous torque $\Lambda\sim r/h \gg 1$. This property of JEDs has two important consequences: (i) the disc itself being weakly dissipative, it may well be unobservable leading to the (wrong) idea that there is no disc; (ii) a JED is cooler than a SAD fed with the same accretion rate, which leads to a smaller aspect ratio $h/r$.

\subsection{Links between jet and disc physics}

The previous sections showed the crucial role played by the magnetic diffusivity within the turbulent JED. On the contrary, jets are best described by an ideal MHD formalism ($\nu_m =\nu_m' =0$)\index{MHD, ideal}.  This leads to the existence of 5 invariants\index{MHD, invariants} along each magnetic surface for polytropic jets\footnote{The magnetic surface rotation rate $\Omega(a)$, the mass to magnetic flux ratio $\eta(a)$, the total specific angular momentum $L(a)$, energy $E(a)$ and entropy $K(a)$.}. The Bernoulli equation\index{MHD equations, Bernoulli} is obtained by projecting the momentum equation~(\ref{eq:2}) along $\vec B_p$ whereas the transfield or Grad-Shafranov equation\index{MHD equations, Grad-Shafranov} by projecting it along $\nabla a$ (perpendicular to $\vec B_p$). For more details see Tsinganos' contribution (this volume). 

MHD simulations\index{wind models, numerical simulations} of jets driven by accretion discs usually assume magnetic field lines rotating at Keplerian speeds and negligible enthalpy leaving therefore 3 free and independent boundary conditions to be specified at each radius (see e.g. \citealt{ande05} and references therein). These are often the density $\rho(r)$, vertical velocity $u_z(r)$ and magnetic field $B_z(r)$ distributions. However, the study of MAES shows that  not all distributions allow for steady state jets: there is a strong interplay between the disc and its jets. Such an interplay appears in the form of analytical links between jet invariants and parameters describing the disc. These links can be found in \citet{cass00a, ferr02} and \citet{ferr04}. Of all disc parameters the disc ejection efficiency\index{disc, ejection index} $\xi$ plays a major role. Indeed, the knowledge of $\xi$ allows to define almost all jet properties. But in order to obtain the allowed values for $\xi$, the full set of MHD equations must be solved.

\section{A glimpse on self-similar solutions}

\subsection{Mathematical method}

This is done by a separation method allowing to
transform the set of partial differential equations (PDE) into two sets of
ordinary differential equations (ODE) with singularities. Now, the gravitational potential in cylindrical coordinates is 
\begin{equation}
\Phi_G(r,z) = - \frac{GM}{r} \left (  1 + \frac{z^2}{r^2} \right )^{-1/2}
\end{equation} 
and it is expected to be the leading energy source and force in accretion
discs. Thus, if JEDs are settled on a large range of radii (so that we do not care about the radial inner and outer boundaries), then the magnetic energy density has to follow gravity in order to match it everywhere. It is therefore justified to look for solutions of the form $A(r,z) = G_A(r) f_A(\frac{z}{r})$ for any physical quantity $A(r,z)$. Moreover, since gravity is a power law of the disc radius, we will use the {\it self-similar} Ansatz\index{self-similarity, radial} $A(r,z)= A_e\left(\frac{r}{r_e}\right)^{\alpha_A}f_A(x)$ where $x = z/h(r)$ is our self-similar variable with $h\propto r$ and $r_e$ is the JED outer radius. Because all quantities have power law dependencies, the resolution of the "radial" set of equations is trivial
and provides algebraic relations between all exponents. The most general
set of radial exponents allowing to take into account {\bf all} terms in
the dynamical equations leads to the following important constraint
\begin{equation}
\beta =  \frac{3}{4}\ +\ \frac{\xi}{2} 
\end{equation}
where the magnetic flux distribution writes $a(r) \propto r^\beta$. As an illustration, the solutions obtained by \citet{blan82} used $\beta=3/4$, ie $\xi =0$. In general, all self-similar models of disc driven jets not addressing the disc dynamics use a magnetic field distribution inconsistent with the jet mass loading \citep{blan82,cont94a,cont94b,ostr97,vlah00}.

All quantities $f_A(x)$ are obtained by solving a system of ODE which can
be put into the form
\[ \left (\begin{array}{ccc} \ldots & & \\ &\bf M &\\ & & \ldots
  \end{array} \right) 
\cdot \left (\begin{array}{c} \frac{df_1}{dx} \\ \vdots \\ \frac{df_n}{dx}
  \end{array} \right)
= \left (\begin{array}{c} \ldots \\ \bf P \\ \ldots \end{array} \right) \]
where ${\bf M}$ is a 8x8 matrix in resistive MHD regime, 6x6 in ideal MHD \citep{ferr95}. A solution is therefore available whenever the matrix ${\bf M}$ is inversible, namely its determinant is
non-zero\index{critical points}. Starting in resistive MHD regime, $det\, {\bf M}=0$ whenever $V^2(V^2 - C_s^2) = 0$
where $C_s$ is the sound speed and $V \equiv \vec u \cdot \vec n$ is the
critical velocity\index{velocity, critical}. The vector $\vec n = (\vec e_z - \frac{z}{r}\vec e_r)(1 + \frac{z^2}{r^2})$ provides the direction of propagation of the only waves consistent with an
axisymmetric, self-similar description (see Tsinganos' contribution). Therefore, close to the disc, the
critical velocity is $V\simeq u_z$, whereas far from the disc it becomes
$V\simeq u_r$ (no critical point in the azimuthal direction). Inside
the resistive disc, the anonalous magnetic resistivity produces such a
dissipation that the magnetic force does not act as a restoring force and the only relevant waves are
sonic. Note also that the equatorial plane where $V=0$ is also a critical
point (of nodal type since all the solutions must pass through it). This
introduces a small difficulty as one must start the integration
slightly above $z=0$. In the ideal MHD region,  $det \, {\bf M}=0$ whenever
$(V^2- V^2_{SM})(V^2-V^2_{FM})(V^2 - V_{An}^2)^2=0$
namely, when the flow velocity $V$ successively reaches the three phase
speeds $V_{SM}$, $V_{An}$ and $V_{FM}$, corresponding respectively to the
slow magnetosonic (SM), Alfv\'en and fast magnetosonic (FM) 
waves. The phase speeds of the two magnetosonic\index{critical points, modified} modes are $V^2_{SM,FM}= \frac{1}{2} \left ( C_s^2 + V_{At}^2 \mp   \sqrt{(C_s^2 + V_{At}^2)^2 - 4C_s^2V_{An}^2}  \right ) $ where $V_{At}$
is the total Alfv\'en speed and $V_{An}= \vec V_{Ap} \cdot \vec n$. These expressions are slightly modified by the self-similar ansatz. Note however that the condition $V=V_{An}$ is equivalent to $u_p=V_{Ap}$.

\begin{figure}
\centering
\includegraphics[height=4cm]{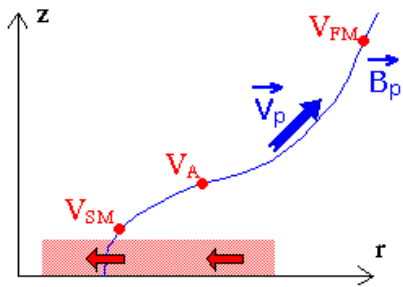}
\caption{Once material has left the resistive MHD zone, it is frozen in a particular field line and encounters the three MHD critical points. The smooth transition between resistive and ideal MHD regimes already selects the MAES parameter space (see Section 2).}
\label{fig:critique}
\end{figure}

How do we proceed~? We fix the values of the four disc parameters $(\varepsilon=h/r, \alpha_m= \nu_m/V_Ah, \chi_m= \nu_m/\nu_m', {\cal P}_m= \nu_v/\nu_m)$ and some guesses for the disc magnetization $\mu$ and ejection efficiency $\xi$. Starting slightly above the disc midplane where all
quantities are now known, we propagate the resistive set of equations using a
Stoer-Burlisch solver for stiff equations. As $x=z/h$ increases, the flow reaches an ideal MHD regime and we shift to the corresponding set of equations. Care must be taken in order to not introduce jumps in the solution while doing so\index{critical points}. The smooth crossing of the SM point can only be done with a critical value for $\mu$. We thus modify our initial guess until the solution gets close enough to the critical point and jump across it (leapfrog method).  The same must be done for the Alfv\'en point which requires a critical value for $\xi$. Each time another guess for $\xi$ is made, one has to find again the corresponding critical value for $\mu$ (Fig.~\ref{fig:critique}). The crossing of the last critical point (FM) does not bring much more information on MAES physics and will be discussed in Section~4.

\subsection{Typical solutions}

Only the most salient features of self-similar accretion-ejection solutions will be discussed here (see \citealt{ferr02} and \citealt{ferr04}).  

\subsubsection{Cold solutions}

Cold solutions are defined here by an isothermal \citep{ferr95,ferr97} or adiabatic \citep{cass00a} energy equation. Since the plasma pressure $P$ is $(h/r)^2$ smaller than the gravitational energy density, such an energy equation ensures that the jet enthalpy is negligible with respect to gravity and magnetic fields \citep{blan82}. 

\begin{figure}
\centering
\includegraphics[height=7cm]{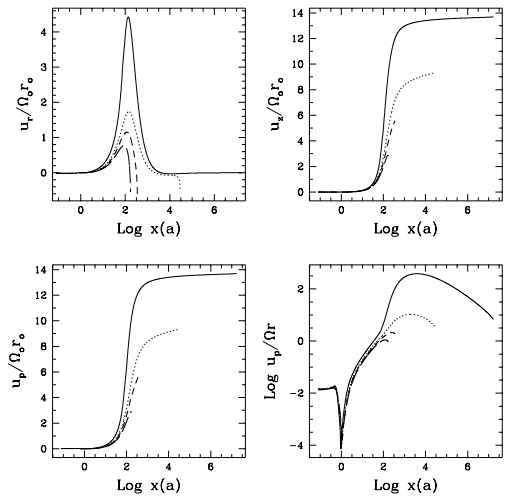}
\caption{Components of the jet poloidal velocity $\vec u_p$ and logarithm of
the ratio of the poloidal to the azimuthal velocity, measured along a
magnetic surface for $\xi= 0.005$ (solid line), 0.01
(dotted line), 0.02 (short-dashed line) and 0.05 (long-dashed line) ($\varepsilon= 10^{-2}$, $\alpha_m=1$). For these typical cold solutions, the jet always reaches its maximum velocity, mainly as a vertical component (the jet opening angle is $\tan \theta= B_r/B_z= u_r/u_z$). Inside the disc, matter is being accreted with a velocity of order $\varepsilon$ the Keplerian velocity $\Omega_o r_o$ \citep{ferr97}.}
\label{fig:cold}
\end{figure}

Figure~\ref{fig:cold} shows the velocity components in both the JED and the jets as a function of the self-similar variable $x$, {\it along a magnetic surface} for typical solutions with $h/r=0.01$ but different ejection efficiencies $\xi$. The disc surface is located at $x=1$ and the Alfv\'en point\index{Alfv\'en, radius} is reached at $x\sim 100$ ($z_A \sim r_A$).  Note that the disc vertical velocity is negative within the disc (material is falling) and becomes positive only slightly before the point where the radial velocity itself becomes positive. This happens roughly at the disc surface (see bottom right panel) but still in the resistive MHD regime. The SM point is crossed at $x\simeq 1.6$. All velocity components are comparable at the Alfv\'en point (this also holds for the magnetic field). Beyond that point, the plasma inertia overcomes the magnetic tension and the magnetic surface opens tremendously. This leads to the build up of a sheared magnetic configuration (the ratio $| B_\phi/B_p |$ increases). Note that the structure of the jet can be characterized by two families of intertwined helices: the plasma streamlines (wound in the same direction as the disc rotation) and the magnetic field lines (wound in the opposite direction). 
 
The magnetic acceleration is so efficient that all available MHD energy is transferred into jet kinetic energy. From Bernoulli equation\index{MHD equations, Bernoulli} one gets analytically the asymptotic\index{velocity, jet asymptotic} jet velocity $v_j = \Omega_o r_o \sqrt{2 \lambda - 3}$ where  $\Omega_o r_o$ is the Keplerian speed at the jet footpoint\index{jet parameters, launching radius} $r_o$ and $\lambda$ is the magnetic\index{jet parameters, magnetic lever arm} lever arm parameter \citep{blan82}. This important jet parameter is actually related to the disc ejection efficiency $\lambda \simeq 1 + 1/2\xi$ \citep{ferr97}.  

\begin{figure}
\centering
\includegraphics[height=4cm]{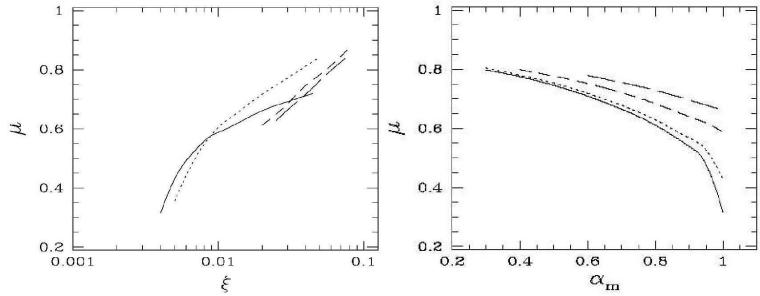}
\caption{Disc parameter space for isothermal jets \citep{ferr97} (adiabatic jets present no qualitative difference, see \citealt{cass00a}). {\bf Left}: disc magnetization $\mu$ as a function of the disc ejection efficiency $\xi$ for $\alpha_m=1$ and various disc aspect ratios: $\varepsilon = h/r= 10^{-1}$ (solid line), $10^{-2}$ (dotted line), $10^{-3}$ (short-dashed line) and $7\ 10^{-4}$ (long-dashed line). The main effect of decreasing $\varepsilon$ is to shift the range of allowed $\xi$ to higher values (but with a more limited range).
{\bf Right}: Influence of the turbulence parameter $\alpha_m$ on the disc magnetization $\mu$ for $\varepsilon= 10^{-1}$ and various ejection efficiencies: $\xi= 0.004$ (solid line), 0.005 (dotted line), 0.01 (short-dashed line) and 0.02 (long-dashed line). The minimum level of MHD turbulence is limited by the value of the induced toroidal field allowing trans-Alfv\'enic jets, whereas the maximum level has been arbitrarily fixed to unity.}
\label{fig:param}
\end{figure}

The disc parameter space has been thoroughly investigated for cold solutions. It is very narrow with typical values $\xi \sim 0.01$ and $0.1 < \mu  < 1$, with the following approximate scaling\index{disc, ejection index}
\begin{equation}
 \xi \sim 0.1\,  \mu^3
\end{equation}
Although its validity holds only in a quite narrow interval, it shows that {\it the stronger the field the more mass is ejected}. No solution has been found outside the range $0.0007 < \varepsilon= h/r < 0.3$ and $0.3 < \alpha_m < 3$ (Fig.\ref{fig:param}). As pointed out previously, there is no solution with a dominant viscous torque.  All solutions exhibit a high degree of collimation: actually, they even undergo recollimation\index{recollimation, shock} towards the axis which should result in a shock \citep{ferr97}. However, the subsequent behaviour of the jet after that shock cannot be treated within self-similarity. 

\subsubsection{Warm solutions}

\begin{figure}[t]
\centering
\includegraphics[height=7cm]{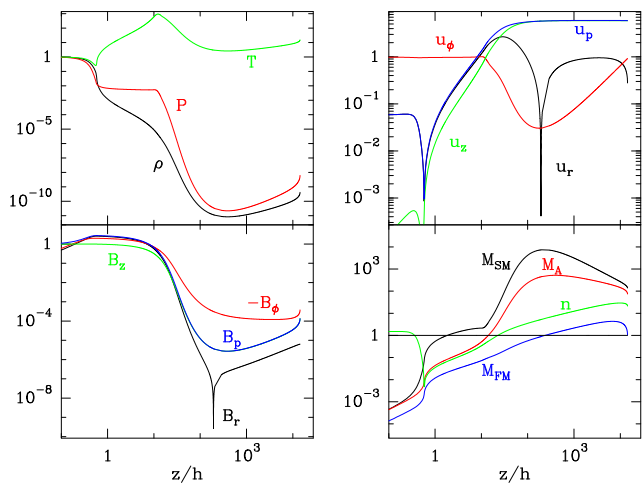}
\caption{Typical super-FM disc wind with $\xi=0.03, \epsilon=
     0.03$ ($h = \epsilon r$). Density, pressure and temperature are
     normalized to their value at the 
     disc midplane, the magnetic field components to $B_z(z=0)$ and 
     the velocities to the Keplerian speed at the anchoring radius
     $r_o$. All magnetic field components remain comparable from
     the disk surface to the Alfv\'en point. Note that the density profile
     inside the disc, where both $u_r$ and $u_z$ are negative, is very
     different from a gaussian. Recollimation\index{recollimation} takes place at $z\simeq 3\ 10^3 r_o$. The lower right panel shows the various critical Mach numbers (e.g. $M_{SM}= V/V_{SM}$) appearing in the self-similar equations. The usual fast Mach number, $n= u_p/u_{FM}$, becomes greater than unity much sooner than the critical one $M_{FM}= V/V_{FM}$ \citep{ferr04}.} 
\label{fig:superFM}
\end{figure}

Warm solutions are obtained by solving Eq.~(\ref{eq:6}) with a prescribed self-similar function $Q$. Several physical effects can be simulated that way:
\begin{itemize}
\item Heat deposition at the disc surface only: the function $Q$ reaches a maximum at the disc upper layers and then decreases rapidly (to recover adiabatic jets). This mimics the effect of disc illumination by stellar UV and X rays. Alternatively, this energy could arise from the dissipation of a small fraction of the accretion energy, released in these layers by turbulence. Remarkably the mass load can be significantly enhanced, with ejection efficiencies up to $\xi \simeq 0.46$ \citep{cass00b}. 

\item Heating of the sub-Alfv\'enic regions: the function $Q$ is non zero in these regions only with subsequent adiabatic or polytropic jets. This mimics the effect of some "coronal" heating as in the solar wind or, alternatively, the pressure due to an inner flow (e.g. stellar or magnetospheric wind) ramming into the disc wind. Under some circumstances, the field lines are forced to open much more than they would which results in a different jet dynamical behaviour. In particular, self-similar jets can smoothly cross the last modified FM critical  point \citep{vlah00,ferr04}. See Fig.~\ref{fig:superFM} for an example.

\item Heating of the whole jet: this has not yet been done in the framework of disc driven jets (but could easily be done by assuming a positive  function $Q$ everywhere in the jet). The reason is that such jets would not be significantly modified by a warmer material (in contrast with stellar winds). However, this is interesting for comparing models to observations. Indeed, observed jets display temperatures of some $10^4$ K that require some heating mechanism(s) overcoming the huge cooling due to the jet expansion (so called adiabatic cooling). It has been shown that ambipolar diffusion\index{ambipolar diffusion} is not enough and that some turbulent\index{heating, turbulent} or shock heating\index{heating, shock} must be at work \citep{garc01b}. 

\end{itemize} 
 
\section{Concluding remarks}

\subsection{What's next~?}

The theory of {\it steady} jet production from Keplerian accretion discs is now completed. The physical conditions required to thermo-magnetically drive jets are known, all relevant physical processes have been included in the framework of mean field dynamics. Of course, there are still many unsolved questions: \\
{\it (i)} Can a sustained MHD turbulence maintain $\alpha_m \sim 1$? This is a huge constraint that deserves a thorough investigation.\\
{\it (ii)} Observations of T~Tauri jets favor solutions with large ejection efficiencies ($\xi \sim 0.1$, \citealt{pese04}) requiring additional heating at the disc surface. A theoretical assessment  of this heating must be undertaken.\\
{\it (iii)} What is the stability of MAES? As will be seen below, there was some claims that MAES were unstable but they were proven to be wrong. On the other hand, jets do show time dependent features and one must clearly go beyond steady state models. On that respect, numerical simulations will be very helpful.\\
{\it (iv)} Disc driven winds do not treat the star-disc interaction. Understanding the whole process of star formation requires now to address this crucial issue as it pinpoints the problem of the stellar angular momentum removal. This is further discussed below.

\subsection{Biases of self-similarity}

Self-similarity allows to take into account all dynamical terms in the equations and, as such, is the best means to solve in a self-consistent way the steady-state accretion-ejection problem. However, there is a price to pay...\index{self-similarity, bias}

{\bf (i) The asymptotic behaviour}\index{asymptotic equilibrium} is obviously biased since, for instance, neither inner nor outer pressures can be taken into account. In fact, no realistic "radial" boundary condition whatsoever can be dealt with. When modeling an astrophysical jet, this implies for instance to truncate the solution at one inner and outer radius. But there is another aspect, less known and more subtle. \citet{cont94a} and \citet{ostr97} obtained jet solutions within the same self-similar framework but with different asymptotic behaviors. The reason stems from the fact that they played around with $\beta$ (flux function $a(r)\propto r^\beta$) as if it were a free function whereas the mathematical matching with a Keplerian disc imposes its value. On the other hand, \citet{pell92} obtained also recollimating\index{recollimation} {\it non self-similar} solutions, which indicates that recollimation can indeed be physical and not entirely due to self-similarity. In fact, it can be shown that recollimation of a jet launched from a Keplerian accretion disc is possible whenever the radial profile of the ejection efficiency $\xi$ is smooth enough \citep{ferr97}. 

{\bf (ii) The regularity conditions} are to be imposed at the modified points\index{critical points, modified} and not at the usual ones (see Tsinganos, this volume). However, these locations coincide for both the slow (SM) and the Alfv\'en points so that one can be confident that there is no bias there. However, this is not so for the fast magnetosonic point. Self-similar trans-FM solutions require an Alfv\'en surface very close to the disc \citep{ferr04}, which can only be done by the action of a large pressure in the sub-Alfv\'enic region. This is obviously a strong bias since it is not clear whether such a pressure is indeed provided in astrophysical objects. Note however that crossing this modified FM point is more a theoretician satisfaction than anything else: it gives no additional physical insight on the disc physics.    

{\bf (iii) The local disc physical conditions} as obtained with self-similar solutions are not biased. The physical processes are well identified and understood and can be sometimes even obtained in a pure analytical manner. They have been also confirmed by numerical experiments of \citet{cass02, cass04, zann04} (although one might object that numerical experiments were actually tested with the help of semi-analytical solutions).   

\subsection{Is accretion-ejection unstable?}

There has been some claims in the literature that the accretion-ejection process itself was unstable\index{instabilities, accretion-ejection} \citep{lubo94b,cao02}. The idea was the following. Start from a steady picture where the accretion velocity $u_r$ at the disc midplane is due to the jet torque. It leads to a bending of the poloidal field lines described by an angle $\theta$ with the vertical. Now imagine a small perturbation $\delta u_r$ enhancing the accretion velocity. Then, according to these authors, the field lines would be more bent ($\theta$ increases) which would lead to lower the altitude of the sonic point. Because the sonic point would be located deeper in the disc atmosphere, where the density is higher, more mass would be henceforth ejected which would then increase the total angular momentum carried away by the jet. This means that the torque due to the jet is enhanced and will, in turn, act to increase the accretion velocity.  Thus, the accretion-ejection process is inherently unstable.

The whole idea of this instability is based upon a crude approximation of the disc vertical equilibrium. In fact, the magnetic field produces a strong vertical compression so that, as $\theta$ is increased, {\it less} mass is being ejected, not more. This has been pointed out by \citet{koni96} and \citet{koni04} and is indeed verified in the full MAES calculations reported here.

\subsection{Magnetic fields in accretion discs}

The necessary condition for launching a self-collimated jet from a Keplerian accretion disc is the presence of a large scale vertical magnetic field close to equipartition\index{magnetic field, equipartition} \citep{ferr95}, namely
\begin{equation}
B_z \simeq  0.2\ \left ( \frac{M}{M_\odot}\right )^{1/4} 
\left ( \frac{\dot M_a}{10^{-7} M_\odot/yr} \right )^{1/2}  \left ( \frac{r_o}{1\mbox{ AU}}\right )^{-5/4 + \xi/2}  \mbox{ G,} 
\end{equation}
This value is far smaller than the one estimated from the interstellar magnetic field (see M\'enard's contribution), assuming either ideal MHD or $B \propto n^{1/2}$ \citep{heil93,basu94,basu95b}. This implies some decoupling between the infalling/accreting material and the magnetic field in order to get rid off this field. This issue is still under debate. The question is therefore whether accretion discs can build up their own large scale magnetic field (dynamo)\index{magnetic field, dynamo} or if they can drag in and amplify the interstellar magnetic field? Although no large scale fields have been provided by a self-consistent disc dynamo, this scenario cannot be excluded. But the latter scenario (advection) seems a bit more natural. 

Let us assume that the disc material is always ionized enough to allow for some coupling with the magnetic field (and use MHD). The outer parts of the accretion disc will probably take the form of a SAD with no jets and almost straight (${\cal R}_m\sim 1$) field lines \citep{lubo94a}. In that case, the steady-state solution of the induction equation for the poloidal field is $B_z \propto r^{-{\cal R}_m}$ \citep{ferr06a}. Hence, as a result of both advection and (turbulent) diffusion, the magnetic field in a SAD will be a power-law of the radius. 

Can a SAD\index{disc, standard} transport $B_z$ and allow for a transition to an inner JED\index{magnetic field, advection}? This will be so if there is some transition radius (the outer JED radius $r_e$) where $\mu= B_z^2/\mu_oP$ becomes of order unity. In a SAD the total pressure writes $P = \frac{\dot M_a \Omega_K^2 h}{6\pi \nu_v} \propto r^{-3/2 - \delta}$ with $h(r)\propto r^\delta$. Since $\delta$ is always close to unity in circumstellar discs, one gets $\mu \propto r^{-\epsilon}$ with $\epsilon \sim 1$. Thus, it can be readily seen that it is indeed reasonable to expect such a transition (computing it is another matter), at least in some objects. The recent Zeeman observation of a magnetic field in the accretion disc of FU Or supports this conclusion \citep{dona05}.  
 
\begin{figure}[t]
\centering
\includegraphics[width=0.9\textwidth]{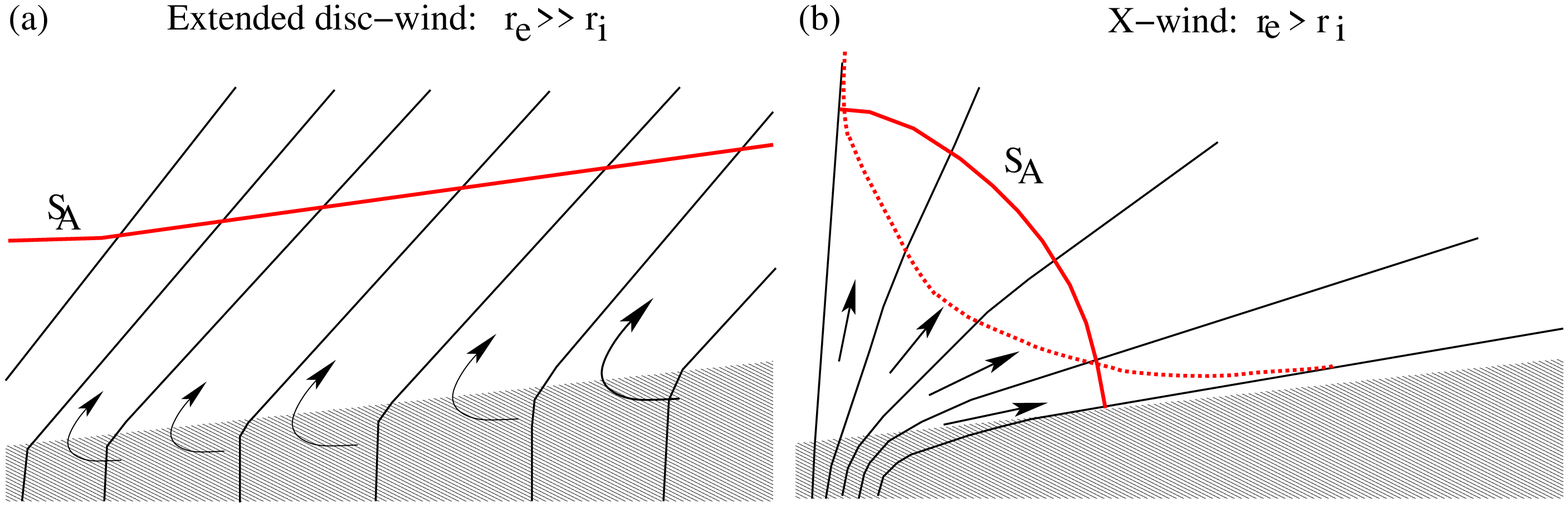}
\caption{Two classes of stationary accretion powered disc winds. {\bf (a)} "extended disc winds", when the magnetic flux threading the disc is large enough so that a large radial extension of the whole accretion disc drives jets ($r_e \gg r_i$). The Alfv\'en surface $S_{\mbox A}$ is expected to adopt a rather conical shape. {\bf (b)} "X-winds", when  the magnetic flux is small and only a tiny disc region is driving jets. The Alfv\'en surface can be either convex or concave, although the latter is probably more physical (since less material can be ejected at the two extremes and the Alfv\'en point is rejected to infinity). Adapted from \citet{ferr06b}.}
\label{fig:winds}
\end{figure}

\subsection{X-winds and disc winds}

The X-wind\index{wind models, disc wind, X-wind} model \citep{shu94a,naji94,shu95,shan98, shan02} is a rich and complex model but, contrary to common belief, it is an accretion-powered wind launched from the accretion disc. In practice, if the amount of magnetic flux threading the disc is large so that $r_e\gg r_i$, then one gets an "extended disc wind", whereas if the magnetic flux is tiny with $r_e \geq r_i$, one gets an "X-wind" (Fig.~\ref{fig:winds}). The dynamics and asymptotic behaviour of jets will differ strongly between an extended disc wind and an X-wind and can thereby be tested against observations \citep{ferr06b}. But this difference arises mainly because of the restricted range in radii in the X-wind case, not because the underlying disc physics is different. The basic phenomena described in Section~2 apply as well for the portion of the disc launching the X-wind. Thus, equipartition fields are required, the "viscous" torque is negligible with respect to the jet torque and the angular momentum carried away by the X-wind is {\it exactly} the same amount lost by the accreting material. As a consequence, X-winds cannot take away any angular momentum from the central star.

Published material on the dynamics of X-winds contains: (i) a scenario for the origin of $B_z$ (stellar) and the star-disc interaction (leading to the opening of some magnetic field lines); (ii) the calculation of the sub-Alfv\'enic ideal MHD jet (elliptic domain defined by prescribed boundary conditions); (iii) a somewhat mysterious "interpolation" to a simple jet asymptotic solution. The following questions remain therefore to be addressed:\\
{\bf (1)} Can the disc afford the imposed mass flux and field geometry? Indeed, the assumed ejection to accretion mass flux ratio of 1/3 from such a tiny region is huge and would require a fantastic ejection efficiency ($\xi$ of order unity or larger). The calculations of JEDs showed that this is unfeasible in a steady way. However, the huge magnetic field gradients required in the X-wind launching region provide a significantly different situation. This has never been analyzed.\\     
{\bf (2)} How good is the transfield equilibrium satisfied? There is no mathematical procedure to find a solution of mixed type (elliptic-hyperbolic) PDEs when the singular surfaces are unknown. The trick used for X-winds provides an incomplete solution, but there is maybe some means to fulfill the transverse equilibrium by using an iterative scheme. In any case, this important point is missing in the current published material.

\subsection{Magnetic star-disc interactions}

Nowadays it seems accepted that a lot if not all young stars have a magnetospheric interaction with their circumstellar accretion disc (see Alencar's contribution, this volume). If one assumes that the disc is threaded by a large scale magnetic field, then the question of how this field is connected to the stellar field arises. First ideas are always simple and so is the stellar magnetic field\index{magnetic field, stellar dipole}, assumed up to now to be dipolar and axisymmetric (see Mohanty et al. 2006). We define here the magnetopause\index{magnetospheric accretion, truncation radius} as the radius 
$r_m$ below which all field lines threading the disc are tied to the star whereas beyond $r_m$, they are disconnected from the star. 

\begin{figure}[t]
\centering
\includegraphics[width=0.9\textwidth]{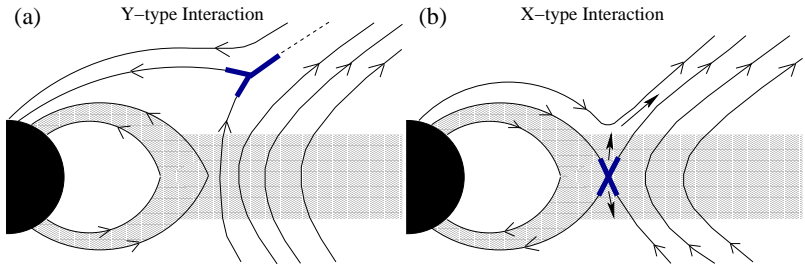}
\caption{Two simple axisymmetric star-disc magnetospheric interactions. {\bf (a)} "Y-type" interaction obtained when the stellar magnetic moment is anti-parallel to the disc magnetic field. A current sheet is formed at the interface between the open stellar field and the disc field. Such a configuration cannot produce per se a wind. {\bf (b)} "X-type" configuration obtained when the stellar magnetic moment is parallel to the disc magnetic field. A magnetic X-point is generated at the disc midplane where the two fields cancel each other. Unsteady ejection ("ReX-winds") can be launched above this reconnection site. Adapted from \citet{ferr06b}.}
\label{fig:magn}
\end{figure}

The case envisioned within the X-wind scenario assumes a stellar magnetic moment anti-parallel to the disc magnetic field \citep{shu94a}. As a consequence, a neutral surface (where $B=0$) appears above each disc surface, illustrated by a limiting poloidal field with a Y shape (Fig.~\ref{fig:magn}). The other case, a stellar magnetic moment parallel to the disc magnetic field, has been proposed by \citet{ferr00}. The two fields then cancel each other at the disc midplane, defining a neutral line\index{magnetic, neutral line} at a radius $r_X$ where reconnection takes place. This configuration gives rise to "Reconnection X-winds" (hereafter ReX-winds) specifically above this zone. 
 
\subsubsection{Accretion curtains}

The first question here is can these simple topologies allow for accretion below\footnote{Accretion is realized beyond $r_m$ by e.g. the jet torque within the JED and, farther away in the SAD,  by the turbulent "viscous" torque.} $r_m$? Disc material\index{magnetospheric accretion, accretion columns} will accrete only if it looses angular momentum and this depends on both turbulence and the magnetic torque due to the magnetosphere. The magnetosphere will try to make the disc corotate with the protostar so the sign of the torque depends on their relative angular velocity. The corotation radius\index{magnetospheric accretion, corotation radius}, $r_{co}= (GM/\Omega_*^2)^{1/3}$, is defined as the radius where the stellar angular velocity $\Omega_*$ is equal to the Keplerian one. This gives an estimate of the real angular velocity of the disc (since the disc magnetic field introduces already a deviation). Roughly speaking, if $r_m> r_{co}$ the star rotates faster than the disc and deposits its angular momentum, whereas if $r_m< r_{co}$, the star rotates slower and thus spins down the disc. Note that $r_m$ denotes roughly the radius where the stellar magnetic field becomes dynamically dominant, namely $\mu >1$. Thus, unless a very efficient turbulent mechanism\footnote{Note that it should be operating when $\mu >1$, while the magneto-rotational instability\index{instabilities, magnetorotational} is already quenched at $\mu \sim 1$ \citep{balb03}.} is operating and transports radially the stellar angular momentum, no accretion is possible when $r_m> r_{co}$ (although such a "propeller" regime is favorable for ejection).

As a consequence, both X-type and Y-type interactions allow for a magnetospheric accretion as long as $r_m< r_{co}$. It is interesting to note that both configurations require an equatorial reconnection zone\index{reconnection} (interesting for sudden energy dissipation and chondrules, \citealt{shu01,goun06}). In the case of a Y-type interaction, it arises because of the requirement that the magnetospheric field makes an angle with the vertical large enough in order to allow the disc material to flow inwards. This assumption implies a magnetic neutral "belt"\index{magnetic, neutral line} at the disc midplane (see Fig.~1 in \citealt{ostr95}), but whose origin and dynamics were not discussed and remain therefore major unsolved issues. In the case of an X-type interaction, the presence of the magnetic neutral line is due to and maintained by the cancellation of the two fields (see fig.\ref{fig:magn}). The accreting disc material can cross the resistive MHD region and is lifted vertically by the strong Lorentz force above the reconnection site. The transition from an accretion disc to accretion curtains can be quite smooth in that case. 

\subsubsection{Stellar spin down}

\begin{figure}[t]
\centering
\includegraphics[height=4cm]{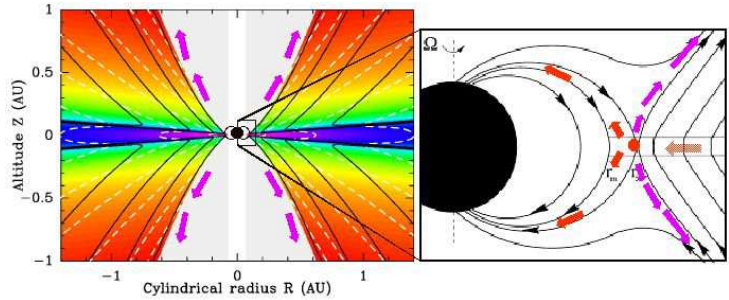}
\caption{The ReX-wind configuration \citep{ferr00}. A MAES is established around a protostar\index{protostar} whose magnetic moment is parallel to the disc magnetic field. This is a natural situation if both fields (disc and stellar) have the same origin. Left: black solid lines are streamlines, white dashed lines are contours of equal total velocity (mainly rotation inside the disc) and the background color scale shows the density stratification. The ReX-wind (arrows) would be confined and channeled by the outer disc wind. Right:   sketch of the magnetic configuration leading to Rex-winds and accretion curtains around the magnetic neutral line at $r_X$. Arrows show the expected time-dependent plasma motion.}    
\label{fig:rex}
\end{figure}

The second question is the issue of the stellar angular momentum removal by winds\index{angular momentum transport, stellar spin down} (see \citet{matt05} for more details and the necessity of winds). As explained earlier, X-winds carry away the angular momentum of the accreting disc material. Thus, such a configuration cannot brake down the protostar (as initially claimed). On the contrary, the X-type configuration provides a very efficient means to do it \citep{ferr00}. The reason is the possibility to launch disc material above the reconnection site. The scenario is the following (Fig.~\ref{fig:rex}). A stationary extended JED is settled in the innermost regions of the accretion disc and provides open magnetic flux to the star. This magnetic field reconnects at $r_X$ with closed stellar field lines: the disc field contributes thereby to transform closed magnetospheric flux into open flux. At the reconnection site, the disc material is lifted vertically and loaded onto these newly opened field lines, tied to the rotating star. Whenever $r_X> r_{co}$, the star is rotating faster than the loaded material and it undergoes a strong magneto-centrifugal acceleration. This gives rise to the so-called ReX-wind\index{wind models, ReX-wind}, whose energy and angular momentum are those of the star. Using a toy-model for the magnetic interaction \citet{ferr00} showed that such winds could brake down a {\it contracting} protostar\index{protostar} on  time scales that are comparable to the duration of the embedded phase (Class 0 and I sources)\index{protostar, Class 0}. The protostar was assumed to rotate initially at breakup speed and, after some $10^5$ to $10^6$ yrs, it has been spun down to 10\% of it despite its contraction and mass accretion.

ReX-winds seem therefore to offer a serious possibility to brake down protostars (to my knowledge, there is no other model in the literature). Note that ReX-winds are probably intermittent by nature because of the unavoidable radial drift of the reconnection site (there is no ejection whenever $r_X< r_{co}$). Dynamically speaking, such an unsteady "wind" should be better described as bullets flowing inside the hollow disc wind. Remarkably the basic features of X-type configurations remain if the stellar dipole is inclined: one would observe in that case precessing bullets channeled by the outer disc wind. Heavy numerical  simulations will be required to test and analyze this scenario.

\end{document}